# Slant-gap plasmonic nanoantenna for optical chirality enhancement

Daniel Lin[†] and Jer-Shing Huang[†,‡,§,*]

[†] *Department of Chemistry, National Tsing Hua University, Hsinchu 30013, Taiwan*

[‡] *Center for Nanotechnology, Materials Sciences, and Microsystems, National Tsing Hua University, Hsinchu 30013, Taiwan*

[§] *Frontier Research Center on Fundamental and Applied Sciences of Matters, National Tsing Hua University, Hsinchu 30013, Taiwan*

**ABSTRACT:** We present a new design of plasmonic nanoantenna with a slant gap for optical chirality engineering. At resonance, the slant gap provides highly enhanced electric field parallel to external magnetic field with a phase delay of $\pi/2$, resulting in enhanced optical chirality. We show by numerical simulations that upon linearly polarized excitation our achiral nanoantenna can generate near field with enhanced optical chirality that can be tuned by the slant angle and resonance condition. Our design can be easily realized and may find applications in circular dichroism enhancement.

Interaction between circularly polarized light (CPL) and chiral matters is of interest because it reveals structural details of molecules that can be critical for their chemical functions.[1] However, such interaction is usually weak due to the mismatch between pitch length of CPL and electronic confinement of the chiral matter.[2,3] One example is the weak circular dichroism (CD) of chiral molecules. CD is the differential absorption of CPL by chiral matters, originating from the coupling between induced electric dipole and magnetic dipole moment.[4] Cohen and co-workers[5] have recently shown that the dissymmetry factor $g$ used for CD is proportional to the optical chirality $C$ [ref. 6] and inversely proportional to the time averaged energy density $U_e$ of the field as

$$g = -K \frac{2C}{\omega U_e} \qquad (1)$$

, where $\omega$ is the angular frequency of the time-harmonic electromagnetic field and $K$ is a factor determined by the electric polarizability and isotropic mixed electric-magnetic dipole polarizability of the chiral matter. Accordingly, they showed enhanced contrast in fluorescence-detected circular dichroism (FDCD) of chiral molecules at energy minima of a standing wave.[7] However, the minimum energy density also means low excitation rate and thus low fluorescence intensity that might limit the detection sensitivity. According to equation (1), an alternative way to enhance CD is to directly enhance the optical chirality $C$ of the field. Since the optical chirality can be expressed as[5]

$$C = -\frac{\varepsilon_0}{2} \omega \text{Im}\{\tilde{E}^* \cdot \tilde{B}\} \qquad (2)$$

, the key to an enhanced CD is to obtain strongly enhanced electric fields that are parallel to the magnetic field with a phase difference of $\pi/2$.[8-10]

Resonant plasmonic nanostructures can localize and enhance optical near field. Therefore, they may function as optical antennas and improve the mismatch between light and nanoscale objects.[11-13] The ability of plasmonic nanostructures to enhance and sculpt optical near fields have recently drawn research attention. On the one hand, various plasmonic nanostructures have been proposed to manipulate the polarization state of light field or to show chiroptical response themselves.[14-26] On the other hand, plasmonic nanostructures and dielectric particles have been reported to create light fields for enhancing chiral light-matter interaction.[8-10,27,28]

Here we present a new design of slant-gap plasmonic nanoantennas and theoretically show that optical chirality of the field in the slant gap is greatly enhanced upon resonant excitation. The working principle is based on the fact that the electric field lines at the vicinity of nanostructures are well perpendicular to the metal boundaries. Therefore, the optical near field in the slant gap automatically provides a component perpendicular to the longitudinal axis of antenna. At longitudinal resonance of the antenna, such electric field component is greatly enhanced in amplitude, delayed $\pi/2$ in phase and well parallel to the magnetic field of external excitation. As a result, an enhanced optical chirality is obtained inside the gap. We choose to excite the slant-gap nanoantenna with the near field of linearly polarized plane wave undergoing total internal reflection (TIR). Such TIR excitation scheme can efficiently suppress the noise from background and is particularly useful for CD analysis of ultra-dilute chiral matters at surface.[29-32] Since the slant-gap nanoantenna is achiral, the nanostructure itself does not give CD signal but creates optical near fields with enhanced OC that facilitates CD analysis of chiral matters.

Figure 1(a) shows the design of the slant-gap plasmonic nanoantenna. The gold nanoantenna has a quadratic cross section (30×30 nm$^2$) and a slant gap (separation in x-direction = 10 nm) tilted in x-z plane by $\alpha$ degree with respect to the surface normal, *i.e.* the z axis. The excitation source is s-polarized plane wave undergoing TIR at air/glass interface on x-y plane. For the excitation configuration in Figure 1(a), the significant near-field components at the TIR interface are

$$\vec{E_x} = \left[\frac{2\cos\theta}{(1-n^2)^{1/2}}\right] A_s e^{-i\delta_s}, \tag{3}$$

$$\vec{H_y} = \left[\frac{2\cos\theta(\sin^2\theta-n^2)^{1/2}}{(1-n^2)^{1/2}}\right] A_s e^{-i(\delta_s-\pi)} \tag{4}$$

and

$$\vec{H_z} = \left[\frac{2\cos\theta\sin\theta}{(1-n^2)^{1/2}}\right] A_s e^{-i\delta_s} \tag{5}$$

, where $\delta_s = \tan^{-1}\left[\frac{(\sin^2\theta-n^2)^{1/2}}{\cos\theta}\right]$, $\theta$ is the TIR incident angle, $n = n_2/n_1 < 1$ and $A_s$ is the amplitude of s-polarized electric field component of the impinging plane waves.[29,31] For a nanoantenna on the interface and aligned in x direction, the $\vec{E_x}$ component of the TIR near field excites the longitudinal surface plasmon resonance and results in highly enhanced optical near field with certain phase shift inside the gap.[13] The field in the gap, $\vec{E_g}$ can be expressed as

$$\vec{E_g} = f_e |\vec{E_x}| e^{-i\varphi_r} \cdot \hat{n} \tag{6}$$

, where $\vec{E_x}$ is the longitudinal electric field component of the TIR near field at the interface, $\varphi_r$ is the phase delay relative to external excitation, $f_e$ is the field amplitude enhancement and $\hat{n}$ is the unit vector normal to the metal surface. At resonance, the phase shift approaches 90 degree and the field enhancement reaches a maximum value. Since the gap is slant, the enhanced field in the gap is no longer parallel to the antenna long axis, as shown in Figure 1(b). As a result, the resonant optical near field automatically provide an out-of-plane electric field component $\vec{E_z}$, which can be expressed as

$$\vec{E_z} = \vec{E_g} \cdot \cos\alpha \tag{7}$$

, with $\alpha$ being the slant angle of the gap relative to the interface (x-y plane). Such an out-of-plane electric field component is parallel to the out-of-plane magnetic field components of the TIR near field ($\vec{H_z}$), leading to an enhanced optical chirality in the gap. Combining equation (2), (3), (5), (6) and (7), the optical chirality can be expressed as

$$C = -\left[\frac{2\varepsilon_0 \omega \cos^2\theta \sin\theta}{(1-n^2)}\right] A_s^2 f_e \cos\alpha \sin\varphi_r \tag{8}$$

For constant $n$, $\theta$ and $\omega$, optical chirality of the near field in the gap is a function of $f_e$, $\varphi_r$, and $\alpha$. In the following, we numerically study the optical chirality in the gap as a function of antenna resonance condition and the slant angle of the gap.

All the simulations in this work are performed using the finite-difference time-domain method (FDTD Solutions v8.6.4, Lumerical Solutions, Inc.). The permittivity of Au are taken from experimental data,[33] and the refractive index of glass is set to be a constant at $n = 1.45$. The mesh size is $2 \times 2 \times 2$ nm³ for the whole slant-gap nanoantenna and an additional mesh override area with mesh size of $1 \times 1 \times 1$ nm³ is used for the gap region. In order to obtain the enhancement effect, we obtain the optical chirality enhancement (OCE) by normalizing the optical chirality at every point in space to the absolute optical chirality of circularly polarized plane wave propagating in vacuum. To estimate the total effect in enhancing CD of randomly distributed chiral matters, we further calculate the overall OCE by integrating OCE over the open space around the antenna. The integration area includes the gap and the antenna vicinity within 10 nm away from the antenna surfaces and excludes the $SiO_2$ half-space and the antenna arms, to which chiral matters are not able to access (Supporting Information). Although the antenna corners also show OCE, the sign alternates and the contribution vanishes when OCE is integrated over space.[9] Therefore, the overall OCE is dominated by the field inside the gap.

Figure 2(a) shows the intensity enhancement and the phase shift of the field inside the gap as a function of the total antenna length. For a fixed excitation frequency at 374 THz (wavelength in vacuum = 800 nm), a total antenna length of 160 nm hits the fundamental resonance and leads to the first maximum of field intensity enhancement. The phase shift is +90 degree with respect to the external excitation field. As the antenna length further increases, higher order resonances gradually emerge.[34] The first higher order resonance appears around antenna length of 480 nm with a phase shift of -90 degree. Compared to the fundamental resonance, the first higher order resonance exhibits lower intensity enhancement and possesses opposite phase shift. This has a direct consequence on the sign and magnitude of the optical chirality of the field in the gap.

The overall OCE as a function of antenna length (Figure 2(b)) reaches a maximum value when the antenna length is at the fundamental resonance. Considering equation (2), it is clear that the maximum OCE value is a direct result of simultaneous fulfillment of maximum electric field and 90 degree phase shift between parallel electric and magnetic field at the fundamental antenna resonance. For antenna lengths that are out of resonance, the overall OCE value quickly drops to zero. As the antenna length hits the first higher order resonance, the overall OCE switches its sign and reaches a local minimum, i.e. a most negative OCE. The absolute OCE value for the higher order resonance is smaller since the field enhancement factor $f_e$ is smaller due to larger plasmon damping on longer antenna arms. As for the opposite sign, it stems from the opposite resonance phase $\varphi_r$ for the first higher order resonance. The phase difference between fundamental and the first higher order resonance is a result of different current distribution in the antenna arms. Figure 2(c) shows the real part of displacement current (Re{$J_x$}) and the out-of-plane electric field component (Re{$\vec{E_z}$}), as well as the OCE distribution of the two resonance conditions on the x-z plane cutting through the antenna center. It can be seen that the current distribution on the antenna arms has opposite sign near the gap for the two resonances. This leads to opposite Re{$\vec{E_z}$} in the gap and therefore opposite sign of OCE in the gap.

Since the magnitude of out-of-plane electric field is proportional to $\cos\alpha$, it is intuitive to choose a small $\alpha$ in order to obtain large OCE. However, varying slant angle can change the antenna geometry significantly and distort the resonance spectrum. Consequently, the field enhancement factor ($f_e$) varies with the slant angle. Therefore, the slant angle must be optimized carefully such that $f_e \times \cos\alpha$ is maximized. Figure 3(a) shows the resonance frequency and the field amplitude enhancement $f_e$ (for 160-nm antenna at 374 THz) as functions of the slant angle. Both quantities vary with the slant angle and the field enhancement $f_e$ reaches the maxima when the resonance frequency matches the desired frequency. As shown in Figure 3(b), increasing the slant angle from 15 to 165 degree, the overall OCE exhibits extreme values with opposite sign at $\alpha = 45$ and $\alpha = 135$ degree and crosses zero value at $\alpha = 90$ degree. The trend of overall OCE obtained from 3-dimensional full-wave FDTD simulations well follows the trend of $f_e \times \cos\alpha$, as shown by the blue dashed line in Figure 3(b). Figure 3(c) shows the OCE distribution for $\alpha = 45$, 90 and 135 degree. It is worth noting that while the sign of OCE in the gap changes with the slant angle, the OCE patterns at antenna outer corners remain the same regardless of the sign of slant slope. This suggests that the field in the gap dominates the overall OCE and the OCE at outer corners vanishes in the integration. In practical applications, changing the sign of the slant slope can be easily achieved by changing the impinging direction of the plane wave in TIR scheme, instead of fabricating new structures. Using the previously proposed stacked antenna[35] may also lead to resonance-enhanced optical chirality in the gap. However, the stacked gap is not open to chiral matters and periodic stacked structures, if possible, would result in cancellation of optical chirality.

In addition to the configurations discussed above, there are other possible excitation geometries, as shown in Figure 4. The field enhancement at the gap center and the overall OCE of the slant-gap antenna at fundamental resonance (antenna length = 160 nm, slant angle = 45 degree) under these illumination geometries are summarized in Table 1. These geometries show either very small or zero overall OCE. For example, normally incident excitation (Figure 4(a)) gives large field enhancement in the gap but results in zero overall OCE because the enhanced electric fields are orthogonal to the magnetic field. As for the excitation geometry shown in Figure 4(b), it is similar to the geometry in Figure 1(a) but using p-polarized incident plane waves. Finite OCE is obtained due to the overlap of in-plane electric ($E_x$) and magnetic field component ($H_x$). However, the OCE is much smaller since it relies on the transverse resonance of the antenna that has relatively low field enhancement at 374 THz. In fact, the slant gap concept is general and the slant angle is not limited to out-of-plane direction (*i.e.* x-z plane in Figure 1). The gap may also be tilted in-plane (*i.e.* x-y plane in Figure 1), as long as the excitation geometry is adjusted accordingly.

In order to maximize the interaction area, the slant-gap concept can be extended into an array of grooves, as depicted in Figure 5(a). The lateral cross section in x-z plane is essentially an array of antenna arms. Similar to solitary gap nanoantennas, the resonance can be tuned by changing the material, the thickness, gap sized and the periodicity. Figure 5(b) and Figure 5(c) show the cross sectional distribution of OCE in x-z plane and x-y plane, respectively. The period of the gold slant groove array is 100 nm. The height and the gap size are 30 nm and 10 nm, respectively. Compared to solitary antennas, the resonance condition changes and the field enhancement decreases due to the inter-arm coupling and the extended dimension in y direction. Nevertheless, the OCE inside the gap has a non-zero value with a sign that is slant angle dependent. The overall OCE obtained from integration over the cross sectional area of a unit cell also shows finite non-zero value that is dominated by the field in the gap. The OCE at outer corners exhibits alternating sign and do not contribute to overall OCE. In practice, depositing a layer of transparent dielectric material may further prevent the chiral matters from interaction with the corner fields. For FDCD detection, such a thin layer can also prevent fluorescence quenching. Slant groove array can be easily fabricated by modern nanofabrication techniques, for example by focused-ion beam (FIB) milling with a tilted angle. Figure 5(d) shows a representative example of an array of slant grooves on a single crystalline gold flake fabricated by FIB milling with a tilt angle of 52 degree relative to the sample surface normal. Other lithography methods such as e-beam lithography might also be applied to fabricate slant grooves by, for example, tilting the sample during the evaporation or dry etching.[36] For chiral matters that give CD response in the UV spectral regime, our concept is applicable but the material should be changed to aluminum.[13,37,38]

In conclusion, we have presented the design of slant-gap nanoantennas excited by total internal reflection scheme. The slant gap offers strong field with enhanced optical chirality that can be controlled and optimized by the resonance condition and the gap slant angle. The slant-gap nanoantenna is achiral and the excitation is linear. Therefore, the nanostructure itself does not give CD response (Supporting Information) but provides optical field with enhanced optical chirality for chiral matters. Together with total internal reflection excitation, the noise from the background can be greatly suppressed. Such enhanced optical chirality by linearly polarized light would be particularly useful for fluorescence-detected circular dichroism analysis of very dilute chiral matters at surface.[39] The enhanced optical chirality may also facilitate photo-induced enantiomeric excess control.[40,41] Our design is simple and easy to realize. We anticipate many applications in plasmon-enhanced chirality detection and control.

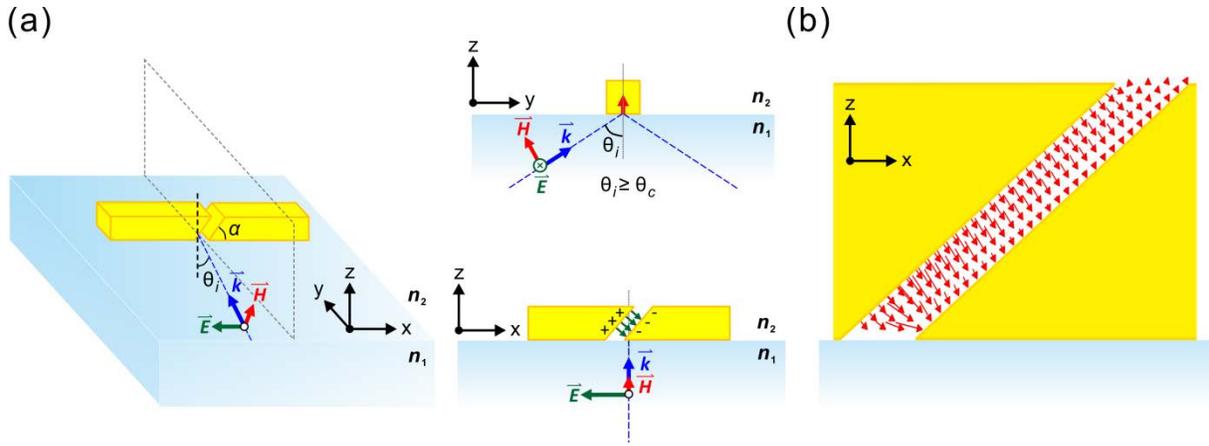

**Figure 1.** (a) Schematic diagrams of a slant-gap plasmonic nanoantenna illuminated with the near field of s-polarized plane waves undergoing total internal reflection. α is the slant angle and $\theta_i$ is the impinging angle of the plane wave and is larger than the critical angle $\theta_c$ for total internal reflection. $n_1$ and $n_2$ are the refractive index of the glass substrate and air, respectively. The electric field $\vec{E}$, magnetic field $\vec{H}$ and the wavevector $\vec{k}$ are indicated with green, red and blue arrows, respectively. (b) The electric field lines (red arrows) in the slant gap are well perpendicular to the metal boundaries.

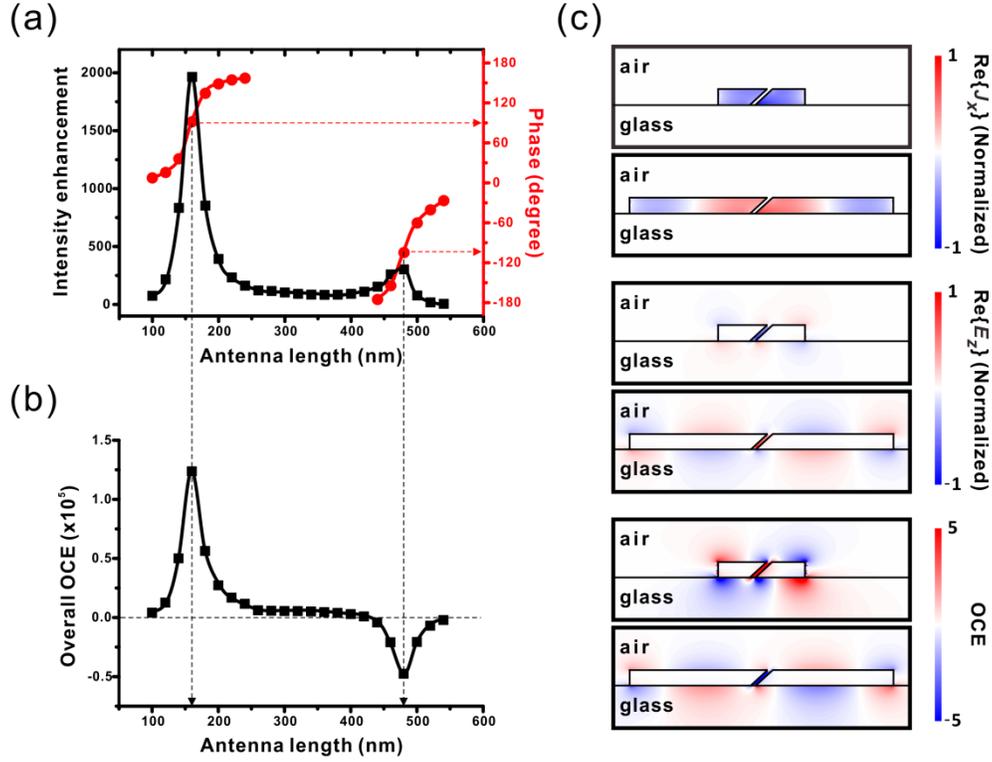

**Figure 2.** (a) Intensity enhancement (black squares) and the spectral phase (red dots) of the field in the slant gap as functions of the total antenna length. (b) Overall optical chirality enhancement (OCE) as a function of antenna length. (c) Distribution of displacement current (top panel), out-of-plane electric-field component ($E_z$, middle panel) and OCE (bottom panel) of the antenna at fundamental resonance (total length = 160 nm) and the first higher order resonance (total length = 480 nm).

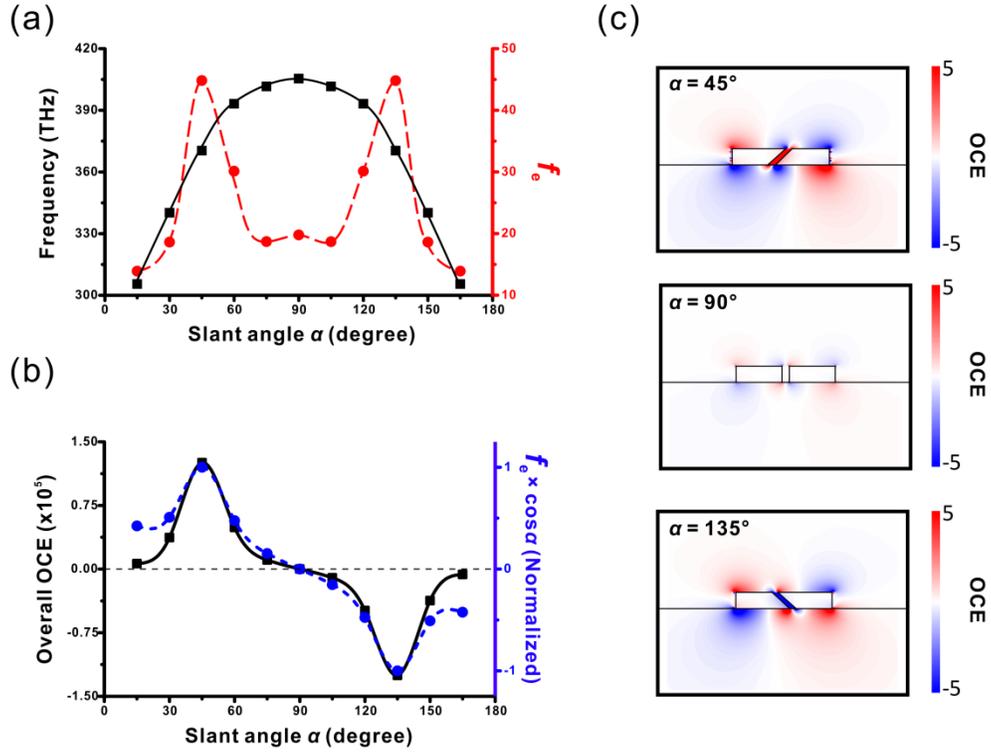

**Figure 3.** (a) Resonant frequency (squares, black solid line) and the field enhancement (dots, red dashed line) as functions of the slant angle of the gap. (b) Overall optical chirality enhancement (OCE, squares, black solid line) and the product of $f_e \times \cos\alpha$ (dots, blue dashed line) as functions of the slant angle of the gap. (c) The cross sectional OCE distributions of various slant angle.

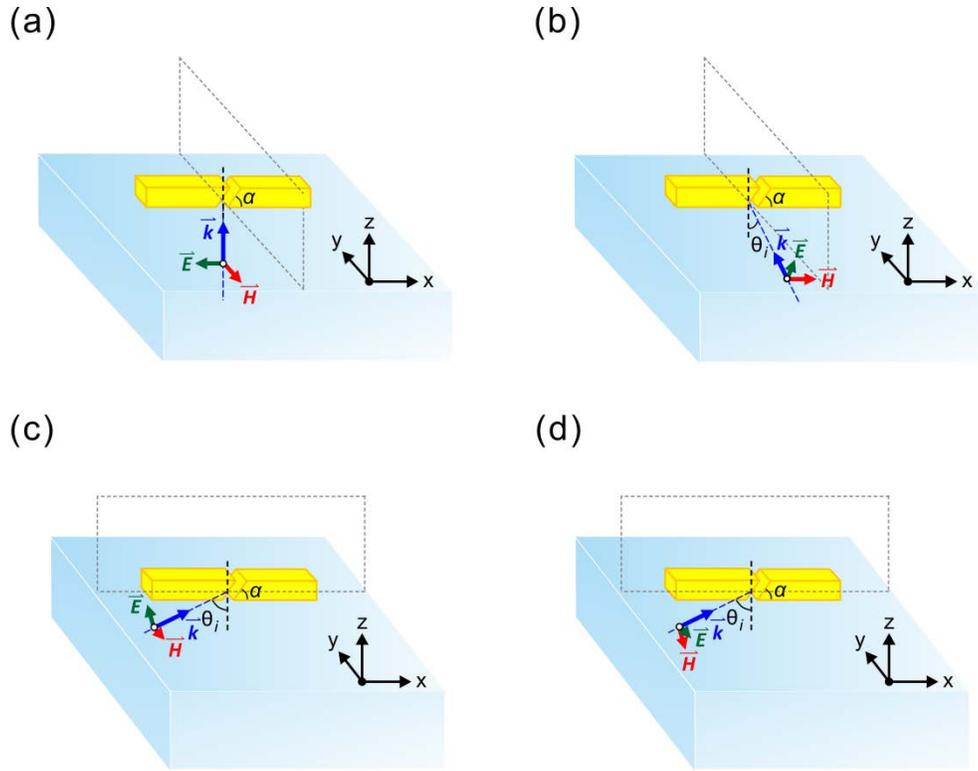

**Figure 4.** Schematic diagram of different excitation configurations, including (a) normally incident excitation linearly polarized in longitudinal antenna axis, (b) p-polarized TIR excitation with magnetic field parallel to the longitudinal axis of antenna, (c) p-polarized TIR excitation with magnetic field perpendicular to the longitudinal axis of antenna and (d) s-polarized TIR excitation with electric field component perpendicular to the longitudinal axis of antenna.

**Table 1.** Gap field enhancement and overall optical chirality enhancement (OCE) of various excitation configurations shown in this work.

| Configuration | Field enhancement ($f_e$) | Overall OCE |
|---|---|---|
| Figure 4(a) | 60.36 | 0 |
| Figure 4(b) | 2.79 | 5340 |
| Figure 4(c) | 28.56 | 0 |
| Figure 4(d) | 0.17 | 0 |
| Figure 1(a) | 44.32 | 123713 |

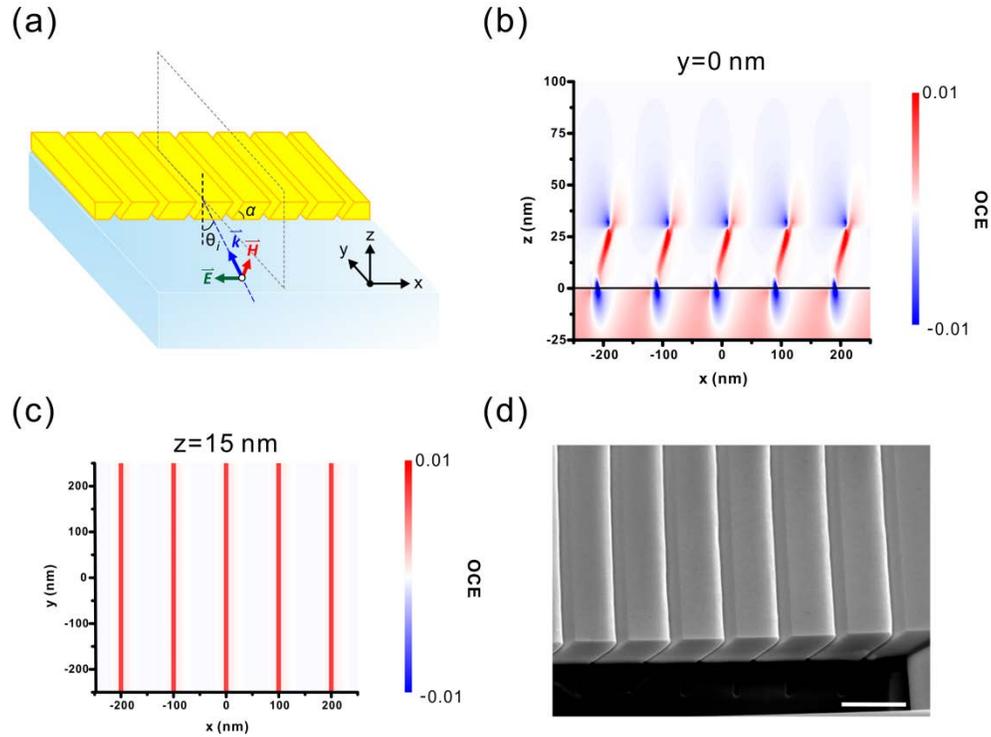

**Figure 5.** (a) Schematic diagram of slant plasmonic grooves illuminated by the near field of s-polarized plane wave undergoing total internal reflection (TIR). The thickness of the gold film is 30 nm and the gap separation in x-direction is 10 nm. $\theta_i$ is the incident angle of the TIR excitation and $\alpha$ is the slant angle of the gap with respect to the surface plane. (b) Cross sectional OCE distribution in x-z plane. (c) OCE distribution in x-y plane cutting through the middle height of the gold film. (d) Representative SEM image of the slant groove array on a single crystalline gold flake. Scale bar is one micrometer.

## ASSOCIATED CONTENT

**Supporting Information**

Additional information about CD of slant-gap nanoantenna and overall OCE calculation are provided. This material is available free of charge via the Internet at http://pubs.acs.org.

## AUTHOR INFORMATION

**Corresponding Author**

*Email: jshuang@mx.nthu.edu.tw

**Notes**

The authors declare no competing financial interests.

## ACKNOWLEDGMENT

Supports from National Science Council of Taiwan under Contract Nos. NSC-99-2113-M-007-020-MY2 and NSC-101-2113-M-007-002-MY2 are gratefully acknowledged. JSH thanks the support from Center for Nanotechnology, Materials Sciences, and Microsystems at National Tsing Hua University.

Supporting Information

# Slant-gap plasmonic nanoantenna for optical chirality enhancement

*Daniel Lin[†] and Jer-Shing Huang[†,‡,§,*]*

[†] Department of Chemistry, National Tsing Hua University, Hsinchu 30013, Taiwan

[‡] Center for Nanotechnology, Materials Sciences, and Microsystems, National Tsing Hua University, Hsinchu 30013, Taiwan

[§] Frontier Research Center on Fundamental and Applied Sciences of Matters, National Tsing Hua University, Hsinchu 30013, Taiwan

*Email: jshuang@mx.nthu.edu.tw

## CD of slant-gap nanoantenna

Figure S1 (a) shows the difference in the absorbance of the slant-gap nanoantenna (total antenna length = 160 nm) illuminated by left-handed and right-handed circularly polarized light. Corresponding dissymmetry factor *g* is plotted in Figure S1 (b). The result shows that the slant-gap nanoantenna is achiral and does not give CD signal.

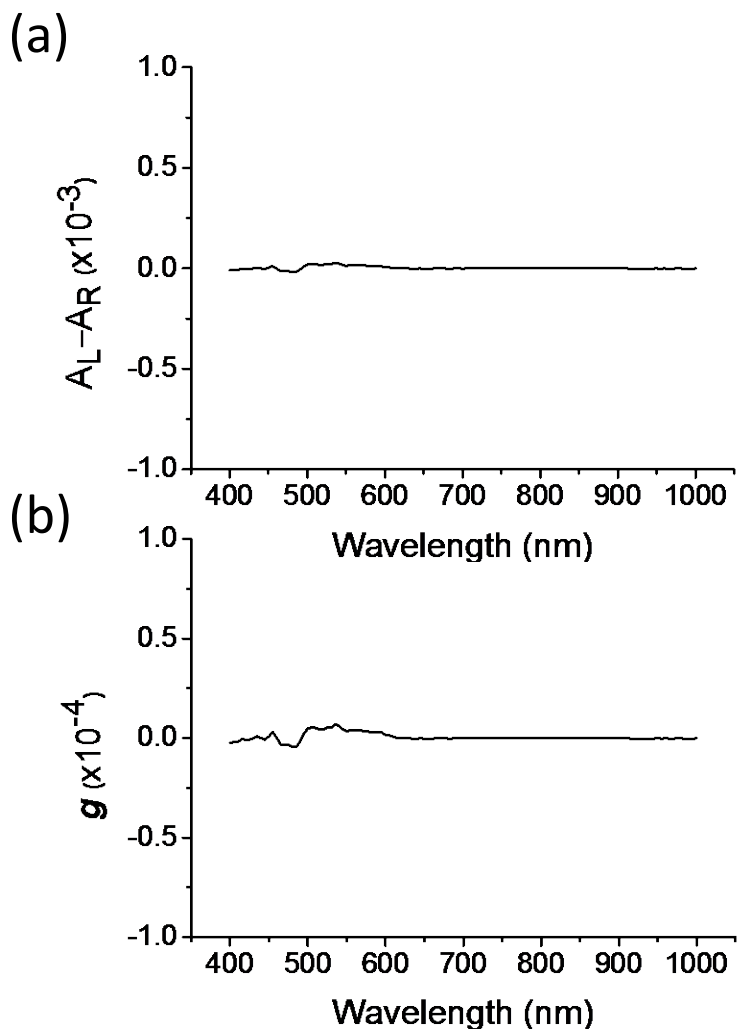

**Figure S1.** Circular dichroism spectrum of the slant-gap nanoantenna. (a) The difference in the absorbance of the slant-gap nanoantenna (total antenna length = 160 nm) illuminated by left-handed and right-handed circularly polarized light. $A_L$ and $A_R$ are the absorbance upon illumination of left-handed and right-handed circularly polarized light, respectively. (b) Corresponding dissymmetry factor *g* as a function of wavelength, calculated using $g = 2\left(\frac{A_L - A_R}{A_L + A_R}\right)$.

**Overall OCE calculation**

    The overall OCE is obtained by integrating OCE value over the vicinity of nanoantenna. Since the near field decays exponentially from the metal surface, we only integrate OCE within a minimum but representative area. The closest distance from the antenna surface to the boundaries of the integration box is set to be a constant $d$, as shown in Figure S2 (a). We exclude the space inside the antenna arms and glass substrate because they are not accessible to chiral matters. Figure S2 (b) shows the overall OCE as a function of $d$ for the 160-nm antenna at the fundamental resonance. It can be seen that the overall OCE saturates at $d = 10$ nm. Therefore we have chosen $d = 10$ nm for our overall OCE calculation.

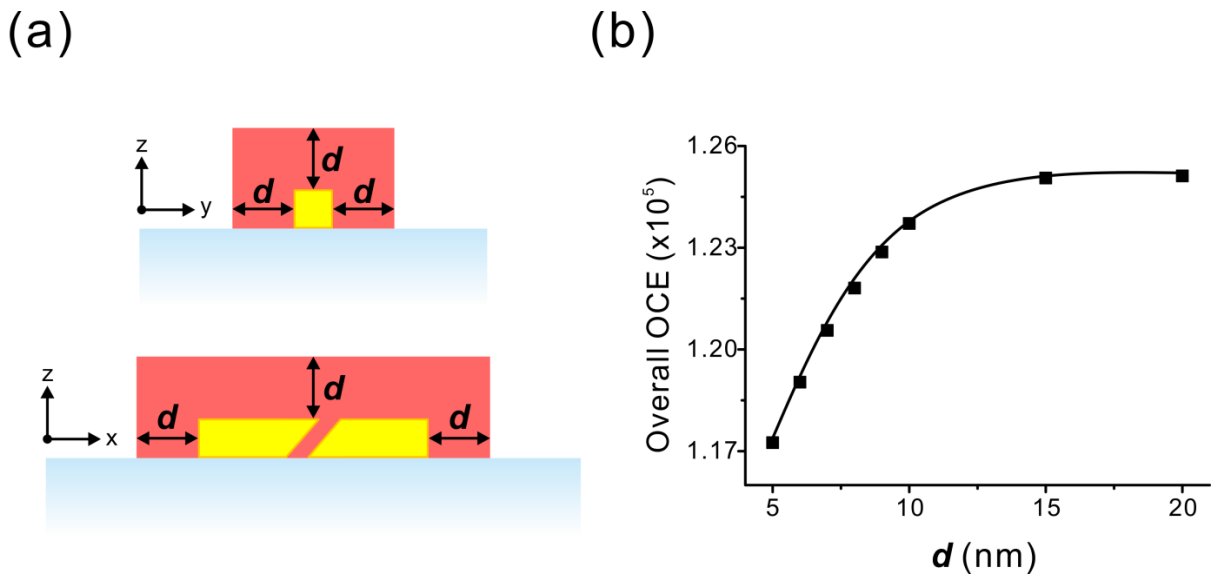

**Figure S2.** (a) The integration area (in red color) for the calculation of overall OCE. The glass half space and the area inside the antenna arms are excluded since they are not accessible to chiral matters. (b) Overall OCE value as a function of $d$ for the 160-nm antenna at the fundamental resonance. Solid line is a guide line for the eye.